\documentclass{phbauth}
\usepackage{graphicx}

\begin{document}

\begin{frontmatter}

\title{
  On the possibility of a superfluid transition in a Fermi-gas of
neutral particles at ultra-low temperatures}

\author[address1]{M.Yu. Kagan\thanksref{thank1}},
\author[address2]{M.A. Baranov},
\author[address2]{Yu. Kagan},
\author[address2]{D.S. Petrov}

\address[address1]{
 P.L.Kapitza Institute for Physical Problems, Moscow 117334, Russia }
\address[address2]{
 Russian Research Center "Kurchatov Institute", Moscow 123182, Russia}

\thanks[thank1]{Corresponding author. Present address:
    P.L.Kapitza Institute for Physical Problems, Kosysin str. 2, Moscow
117334, Russia. E-mail: kagan@kapitza.ras.ru}

\begin{abstract} We predict a possibility of triplet Cooper pairing in
a Fermi-gas of neutral particles in a confined geometry of magnetic
traps at ultra-low temperatures. We evaluate a superfluid transition
temperature and analyze the difference between pairing in free space
and in confined geometry. We also consider in details a case of
fermionic $^6$Li.
\end{abstract}


\end{frontmatter}

One of the most important events of the past
several years in condensed matter physics was the discovery of
Bose-Einstein condensation of the alkali elements
$^{87}$Rb \cite{1}, $^7$Li \cite{2} and $^{23}$Na \cite{3} in confined
geometry of magnetooptical traps. In the present paper we consider the
fundamental possibility of achieving superfluid instability of a
different type --- with respect to Cooper pair formation in a non-ideal
atomic Fermi-gas with a large scattering length. We shall consider both
the case of an attractive scattering length
$a<0$ (for $^6$Li $a=-2.3\cdot 10^3\ {\rm\AA}<0$ \cite{4}) and a
repulsive case
$a>0$.

We consider a multi-component Fermi-gas \cite{5} with
short range interaction and equal mases of each component. For a
Fermi-gas in a magnetic trap each component has a hyperfine origin and
corresponds to different
$J_z$-projection of nuclear spin $\hbox{\bf J}$. Hence the number of
the components $\nu=2J+1$. Note that for $^6$Li $J=1$ and thus $\nu=3$.

For an attractive scattering length $a<0$ the pairing occurs in the
$s$-wave channel ($l=0$) \cite{6}. According to the Pauli principle,
Cooper pairs in this case can only be formed by the atoms which are in
different hyperfine states. Therefore the critical temperature $T_c$ of
the transition is very sensitive to the difference in the concentrations
of the two hyperfine components. When the densities are identically
equal, the standard BCS mechanism of Cooper pairing yields \cite{6}:
\begin{equation}
  T_{c0}=0.28 \varepsilon_F \exp \left( -{\lambda}^{-1}\right),
  \end{equation}  where $\lambda={2a|p_F|}/{\pi}$ is
the gas parameter.

However, already for ${\Delta n}/{n}>{T_{c0}}/{\varepsilon_F}$ the
$s$-wave pairing is completely suppressed. For $^6$Li one has $T_{c0}
\approx 30$ nK for an atom density $n=4
\cdot 10^{-12}\ {\rm cm^{-3}}$ \cite{6} and the existence of $s$-wave
pairing requires
\cite{8} ${\Delta n}/{n}<3\cdot 10^{-2}$.

For positive scattering length ($a>0$) the $s$-wave pairing is
completely impossible. However, even in this case Fermi gas becomes
superfluid at a low enough temperatures due to Kohn-Luttinger mechanism
of Cooper pairing
\cite{9}.  This mechanism gives rise to an unconventional
$p$-wave pairing, analogous to A$_1$ phase of superfluid $^3$He \cite{5}
($l=1$,
$J_{1z}=J_{2z}$).  The Pauli principle in this case allows us to have
Cooper pairs formed by two particles that are in the same hyperfine
states, whereas the particles in other hyperfine states participate
only in the formation of the effective pairing interaction.  To be more
specific, the effective interaction is given by a sum of loop diagrams
\begin{equation} V_{\mathrm{eff}}(q_1)=(-1)\left( \frac{4\pi
a}{m}\right)^2
\sum_{j=2}^{\nu} \Pi_{jj}(q_1), \end{equation} where \[
  \Pi_{jj}(q_1)=\int \frac{\mathrm{d}^3 p}{(2\pi^3)} \frac{n_j(p)-
  n_j(p+q_1)}{\varepsilon_j(p) - \varepsilon_j(p+q_1)} \] is the
  polarization operator for the $j$-th component. When the densities of
the $\nu$ components are close to each other $n_1 \approx n_2 \approx
\ldots \approx n_{\nu}$, the critical temperature of $p$-pairing is the
same for each component and does not depend upon the sign of $a$:
\begin{equation} T_{c1}\sim
  \varepsilon_{F1} \exp \left\{ -\frac{13}{(\nu-1) \lambda_1^2}
\right\},
  \end{equation}  where $\lambda_1={2ap_{F1}}/{\pi}$.

For $^6$Li the $p$-wave critical temperature
$T_{c1}\sim 10^{-8}$ K for densities $n_1 \approx n_2 \approx n_3 \sim
10^{13}\ {\rm cm^{-3}}$ adjusted with the accuracy
${\Delta n}/{n}\sim 0.1$. To increase the $p$-wave critical temperature
further we must exploit a strongly non-monotonic dependence of $T_{c1}$
upon the density ratio
${n_1}/{n_j}$. In paper \cite{5} it was predicted that this
dependence has a pronounced and extended maximum at
${n_1}/{n_j}\approx 3$ (${p_{F1}}/{p_{Fj}}\approx
1.4$). In an optimal situation when $n_1 \approx 3n_2 \approx \ldots
\approx 3n_{\nu}$ a critical temperature of the component 1 reads:
\begin{equation}
  T_{c1}=\varepsilon_{F1} \exp \left\{-\frac{7}{(\nu-1)
  \lambda^2_{\mathrm{eff}}} \right\}, \end{equation} where
$\lambda^2_{\mathrm{eff}}=\left( {2a}/{\pi} \right)^2 p_{F1}
p_{Fj}$.

For $^6$Li this formula yields $T_{c1} \sim 10^{-7}$ K for $n_1\sim
10^{14}\ {\rm cm^{-3}}$. Note that even for these high densities
$\lambda_{\mathrm{eff}}{\ \lower-1.2pt\vbox{\hbox{\rlap{$
        <
        $}\lower5pt\vbox{\hbox{$\sim$}}}}\ } 1$ and we can still
satisfy stability conditions for an attractive case.

All our results are obtained in a spatially
homogeneous situation. In a confined geometry of a magnetic trap we
must satisfy a condition
$\xi<R_{\varepsilon_F}$. This condition means that the size of Cooper
pair
$\xi={\hbar v_F}/{T_c}$ is much smaller than the
characteristic size of a trap. For a parabolic trap
\cite{8} $R_{\varepsilon_F}={v_F}/{\Omega}$, where
$\Omega$ is a level spacing. Hence, our results are valid in a
semi-classical limit where
$\Omega<T_c<\varepsilon_F$.

 \begin{ack}
   M.Yu.K. acknowledges financial support of President Eltzin's grant
 96-15-96942.
  \end{ack}

\end{document}